\documentclass[12pt]{article}
\usepackage{graphicx}
\usepackage{amsmath,amssymb}
\newcommand{\beq}{\begin{eqnarray}}
\newcommand{\eeq}{\end{eqnarray}}

\def\gsim{\displaystyle\mathop{>}_{\sim}}
\def\lsim{\displaystyle\mathop{<}_{\sim}}

\begin{document}

\begin{center}
{\large  \bf
Relation between the separable and one-boson-exchange potential 
for the covariant Bethe-Salpeter equation}
 
 \vspace*{1cm}
Y. Manabe, A. Hosaka and H. Toki\\
\vspace*{2mm}
 {\it Research Center for Nuclear Physics (RCNP), Osaka University\\
 Ibaraki, Osaka 567-0047, Japan}\\
\end{center}
\vspace*{0.2cm}

\abstract{
 We investigate the relation between the rank I separable potential for the
 covariant Bethe-Salpeter equation and the one-boson-exchange potential.
After several trials of the parameter choices,
it turns out that it is not always possible to reproduce the phase-shifts
  calculated
 from a single term of the one-boson-exchange potential especially of the 
$\sigma$-exchange term, separately by the
 rank I separable potential.
Instead, it is shown that the separable potential is useful to parameterize the
 total nucleon-nucleon interaction.
}

\vspace*{1cm}

\section{Introduction}

Relativistic approaches of the nuclear physics becomes 
important for high momentum phenomena, in particular 
for those associated with spin observables~{\cite {sabuandtoki}}.  
Furthermore, the phenomena where pions appear as in many nuclear physics
phenomena,
the relativistic treatment is essential, since the fundamental 
chiral symmetry is related to the relativistic nature of 
particles~{\cite {qbacs}}.  

The starting point of the relativistically covariant 
theory is the Bethe-Salpeter equation for the two 
nucleon system~{\cite {bsgeneral}}.  
In order to overcome a difficulty of solving 
the  integral equation, a separable 
interaction is often employed, primarily as a mathematical 
manipulation~{\cite{yam}}.  
Ignoring a possible dependence on the total momentum 
$P = p_1 + p_2 = p_1^\prime + p_2^\prime$, 
where $p_1$ and $p_2$ are the momenta for the initial two 
nucleons, while $p_1^\prime$ and $p_2^\prime$ for the 
final state ones, 
the interaction is given as a function of the 
relative momenta
$p = (p_1 - p_2)/2$, 
$p^\prime = (p_1^\prime - p_2^\prime)/2$, 
\beq
V_{\rm sep}(p^\prime, p) = \lambda g(p^\prime)g(p) \, .
\label{Vsep}
\eeq
This rank I separable potential is non-local and is very much 
different from the widely used one-boson-exchange 
potential (OBEP) which is, to the leading order, given as
a function of the momentum transfer $ q = p^\prime - p$ 
and is local:
\beq
V_{\rm OBEP}(p^\prime, p) = V(q)\, .
\label{Vobep}
\eeq
In their form of Eq.~(\ref{Vsep}) and (\ref{Vobep}), they can not be 
equivalent, but, instead, higher rank separable interactions 
may be used to generate the OBEP when infinitely many terms are 
introduced.  
Practically, a finite rank (usually up to rank three)  potential 
is used with a finite number of parameters determined in the 
phase shift analysis~{\cite {Plessas:ty}}.  

In this paper, we investigate whether the parameters of the 
separable potential may be related to those of the 
OBEP, since the latter is considered to be physically 
more fundamental, at least for longer range part of the 
NN interaction.  
By doing this, we expect that the separable potential 
can be understood with physics ground, not just a mathematically 
convenient tool.

In section 2, we briefly formulate the Bethe-Salpeter 
equation and provide an analytic solution when a rank I separable 
potential is used for a single channel problem of $l = 0$.  
The rank I separable potential and OBEP are then related in the 
long wave length approximation.  
In section 3, we compare the phase shifts calculated from 
the two potentials.  
Final section is devoted to conclusions of the present work.  

\section{The BS equation with a separable interaction}

Let us consider a single channel equation for $^1S_0$.
This is sufficient for our present qualitative discussions.  
After angular integration, the BS equation is given by~{\cite {Bondarenko:2002zz}} 
\beq
T(p^\prime, p; s) = V(p^\prime, p)
+ \frac{i}{4\pi^3} \int dk_0 k^2 dk
\frac{V(p^\prime, k) T(k, p; s)}
{\left( \frac{\sqrt{s}}{2} - e_k + i\epsilon \right)^2
-k_0^2 } \, , 
\label{bseq}
\eeq
where $T(p^\prime, p; s)$ is the $T$-matrix, 
$V( p^\prime, p)$ the interaction kernel, 
$s = (p_1+p_2)^2$ and 
$e_k = \sqrt{\vec k^2 + M_N^2}$
with $M_N$ being the mass of the nucleon~\cite{Bondarenko:2002zz}.  
The momentum variables are for four momentum, e.g., 
$p =(p_0, \vec p)$, etc.  
In Eq.~(\ref{bseq}), we have considered the equation in the 
center of mass system.  
The rank I separable ansatz assumes to write the interaction 
\beq
V_{\rm sep}( p^\prime, p) =
\lambda g(p^\prime)g(p)
\eeq
with a coupling constant $\lambda$ and a function $g(p)$ as a scalar function
 of
$p$ and $p{}^{\prime}$.  
The $T$-matrix is then obtained in the separable form as 
\beq
T(p^\prime, p; s) &=& \tau (s) g(p^\prime)g(p) \, , 
\nonumber \\
\tau (s) &=& \frac{1}{\frac{1}{\lambda}+h(s)}\, ,
\label{tau}
\eeq
where 
\beq
h(s) = - \frac{i}{4\pi^3}
\int dk_0 k^2 dk 
\frac{g(k)^2}{\left( \frac{\sqrt{s}}{2} - e_k 
+ i\epsilon \right)^2 - k_0^2}\, .
\label{hfunction}
\eeq
Phase shifts are then given by the relation
\beq
T(p^\prime, p; s) = - \frac{16\pi}
{\sqrt{s} \sqrt{s-4{M_N}^2}} e^{i\delta} \sin \delta \, , 
\label{t_and_delta}
\eeq
where the relative momenta are 
$p = (0, \vec p)$, $p^\prime = (0, \vec p^{~\prime})$ 
for on-shell nucleons.  
 Finally the phase shift $\delta(s)$ can be presented by 
\beq
\cot\delta(s) = - \frac{\lambda^{-1}+{\rm Re}(h(s))}{{\rm Im}(h(s))} \, . 
\label{phaseshift}
\eeq

Now important terms of the OBEP can be written as 
\beq
V_{\sigma}(q) = -g_{\sigma}^2 \frac{1}{-q^2+m_{\sigma}^2}
\Bigl(\frac{\Lambda_{\sigma}^2-m_{\sigma}^2}
{\Lambda_{\sigma}^2-q^2}\Bigr)^2 \, , 
 ~V_{\omega}(q) = g_{\omega}^2 \frac{1}{-q^2+m_{\omega}^2}
\Bigl(\frac{\Lambda_{\omega}^2-m_{\omega}^2}
{\Lambda_{\omega}^2-q^2}\Bigr)^2\, , 
\label{Vsigmaomega}
\eeq
\beq
V_{\pi}(q) = \frac{g_{\pi}^2}{4M_N^2} \frac{ q^2}{-q^2+m_{\pi}^2}
\Bigl(\frac{\Lambda_{\pi}^2-m_{\pi}^2}
{\Lambda_{\pi}^2-q^2}\Bigr)^2\, , 
~V_{\rho}(q) = \frac{g_{\rho}^2}{2M_N^2} \frac{ q^2}
{-q^2+m_{\rho}^2}
\Bigl(\frac{\Lambda_{\rho}^2-m_{\rho}^2}
{\Lambda_{\rho}^2-q^2}\Bigr)^2\, ,
\label{Vpirho}
\eeq
where $q =p-p^\prime=  (0, \vec p-\vec p^{~\prime})$.
Here the masses $m_{\alpha}$, the coupling constants $g_{\alpha}$ and the
 cutoff parameters $\Lambda_{\alpha}(\alpha=\sigma,\omega,\pi,\rho)$ are given
in  Ref.~{\cite {bonn}}, and are summarized in Table~\ref{table1}.
In Eqs.~(\ref{Vsigmaomega}) and (\ref{Vpirho}) we picked up the dominant
 piece of the one boson exchange potential.
The higher order terms are proportional to the initial and final relative
 momenta $p$ and $p'$.
For the $\rho$-exchange potential, we use only the tensor coupling term,
where the correction from the vector term is about 5 $\%$.
The coupling strength given in Table~\ref{table1}(in the second row)
produces only the $f$-coupling in  Ref.~{\cite {bonn}}.

For the parameterization of the separable potential, 
we assume the Yukawa function for $g(p)$ with the same 
mass parameter $m$ as in the OBEP, 
$g(p) = 1/(p^2 - m_b^2)$.  
Then we try to impose that $V_{\rm sep}$ equals $V_b$ in the long 
wave length limit~{\cite {Bondarenko:2002zz}}{\footnote {The relations shown
 here differ from those of Ref~{\cite {Bondarenko:2002zz}}, where extra factor 
of $4\pi^2$ was erroneously included. }:
\beq
V_{\rm sep}(0,0) = V_b(0,0) \, ,
\label{comparison}
\eeq
which determines the strength $\lambda$.  
In this way, we have a separable potential approximately 
related to the OBEP
\beq
V_{\rm sep}(p^\prime,p) =   \lambda_b
\frac{1}{p^{\prime 2} - m_b^2}  \frac{1}{p^2 - m_b^2} \, ,
\eeq
where $\lambda_b$ are given by
\beq
\lambda_{\sigma}=-g_{\sigma}^2 m_{\sigma}^2
\Bigl(1-\frac{m_{\sigma}^2}{\Lambda_{\sigma}^2} \Bigr)^2\, ,
\label{Vsigma}
~\lambda_{\omega}= g_{\omega}^2 m_{\omega}^2
\Bigl(1-\frac{m_{\omega}^2}{\Lambda_{\omega}^2} \Bigr)^2\, ,
\label{Vomega}
\eeq
\beq
\lambda_{\pi}= -\frac{g_{\pi}^2 m_{\pi}^4}{4 M_N^2}
\Bigl(1-\frac{m_{\pi}^2}{\Lambda_{\pi}^2} \Bigr)^2\, ,
\label{Vpi}
~\lambda_{\rho}= -\frac{g_{\rho}^2 m_{\rho}^4}{2 M_N^2}
\Bigl(1-\frac{m_{\rho}^2}{\Lambda_{\rho}^2} \Bigr)^2\, .
\label{Vrho}
\eeq
In the case of $\pi$ and $\rho$, we excluded $q^2$ dependence in the
numerator of the Eqs.~(\ref{Vpirho2}), otherwise we can not
 determine the $\lambda$ parameter.
It  corresponds to excluding the $\delta$-function term in the $r$ space, 
namely for the Eq.~(\ref{comparison}) we have used 
\beq
V_{\pi}(q) = -\frac{g_{\pi}^2}{4M_N^2} \frac{m_{\pi}^2}{-q^2+m_{\pi}^2}
\Bigl(\frac{\Lambda_{\pi}^2-m_{\pi}^2}
{\Lambda_{\pi}^2-q^2}\Bigr)^2\, , 
~V_{\rho}(q) = -\frac{g_{\rho}^2}{2M_N^2} \frac{m_{\rho}^2}
{-q^2+m_{\rho}^2}
\Bigl(\frac{\Lambda_{\rho}^2-m_{\rho}^2}
{\Lambda_{\rho}^2-q^2}\Bigr)^2\, .
\label{Vpirho2}
\eeq  
The numerical values of the $\lambda'$s are also given in the
 Table~\ref{table1} (last column).

\section{Comparison of phase shifts}
 We have calculated phase shifts using the BS Eqs.~(\ref{bseq}) --
 (\ref{hfunction}), which are compared 
with those obtained from the OBEP.
Since we make the comparison at relatively low energy region,
it is sufficient to solve the Schr\"odinger equation. 
Here we compare  various phase shifts calculated by using a single term
 corresponding to 
$\sigma$-, $\omega$-, $\pi$- or $\rho$-exchange potentials. 
The resulting phase shifts are shown in Figs.~\ref{fig:bonnandsp}.
For later use, we show the one boson exchange potential of the ${}^1S_0$
 channel as a function of $r$ in  Figs.~\ref{fig:potential},
where various terms of the OBEP are shown separately.
The thick solid line in Fig.~\ref{fig:potential}-(a) is the total potential
including the $\sigma$, $\omega$, $\pi$ and $\rho$ exchange potentials   

Now we discuss the phase shifts calculated from each meson exchange potential.
\begin{itemize}

\item

Fig.~\ref{fig:bonnandsp}-(a) shows the phase shifts calculated from the
 potentials of the $\sigma$ channel as functions of $T_{lab}$, the kinetic
 energy in the laboratory frame. 
Here $T_{lab}$ is related to $s$ by
\beq
T_{lab}=\frac{s-4M_N^2}{2M_N}\, .
\label{sandtlab}
\eeq 
The thick solid line represents the phase shift for the separable potential,
 and the thin solid line for the OBEP.
As shown in Fig.~\ref{fig:bonnandsp}-(a), both phase shifts start from 180
 degrees, indicating a strong attraction as accommodating one bound state.
Indeed, as shown in Fig.~\ref{fig:potential}-(a), the depth of the
 $\sigma$-exchange potential of the OBEP reaches about 200 MeV at 0.75 fm.
The strong attraction of the OBEP causes the raising behavior at $T_{lab=0}$, 
which turns to decreasing at $T_{lab}=20$ MeV.
On the other hand, the separable potential can not be that strongly attractive.
This can be checked by analyzing the scattering matrix of Eq.~(\ref{tau}),
which will be discussed later.
In fact, the phase shift approaches the upper limit as indicated by the dashed
 line in Fig.~\ref{fig:bonnandsp}-(a) in the limit $\lambda\to-\infty$.
Furthermore we have checked that it is not possible to reproduce the  strong
 attraction of the $\sigma$-exchange potential, whatever $m_b$ value of the
 separable potential we choose.
The fact that the separable potential can not be too strong has been discussed
previously~{\cite {newton}}. 
\item
 
Fig.~\ref{fig:bonnandsp}-(b)  shows the phase shifts calculated
 from the potentials of the $\omega$ channel as functions of $T_{lab}$.
The phase shifts calculated from the two interactions  (Separable and OBEP)
show repulsive nature
as it starts from 0 degree and decreases as the energy increases. 
As shown in Fig.~\ref{fig:potential}-(a), the repulsive force of the
 $\omega$-exchange potential is very strong.
As shown in Fig.~\ref{fig:bonnandsp}-(b), the result of the separable potential
 resembles that of OBEP.
However, that strong repulsion of the OBEP can not be reproduced by
the separable potential, which is similar to the case of the $\sigma$-exchange
 potential.
Once again, as shown by the dashed line in Fig.~\ref{fig:bonnandsp}-(b)
 there  is a lower bound of the phase shift of the separable potential in the
 limit
$\lambda\to\infty$.     
However if we allow $m_b$ to change, it is possible to reproduce the phase
 shift of the $\omega$-exchange potential by using a parameter set of, for
 instance,
 $m_b=630$ MeV and $\lambda=81.5\times 10^6$ MeV${}^2$.
To make $m_b$ small corresponds to the increase of repulsion. 
At this point, we recognize that the physical meaning of the mass parameter
$m_b$ in the separable potential is different from that in the OBEP.

\item
Fig.~\ref{fig:bonnandsp}-(c)  shows  the phase shifts calculated
 from the potentials of the $\pi$ channel as functions of $T_{lab}$.
The phase shifts calculated from the two interactions look very different.
Fig.~\ref{fig:potential}-(b) shows that  $\pi$-exchange potential is
 attractive  at  long distances $r \gsim $ 1 fm and repulsive at middle and 
short distances $r \lsim$ 1 fm.
Therefore the phase shift calculated from OBEP starts from 0 degree, raising
 at  first, then turns to  decrease at  $T_{lab}\sim 5$ MeV and  becomes
 repulsive at  $T_{lab}\sim 20$ MeV.
Due to the form factor, the one-pion-term of the OBEP here is written as a sum
of the long range attraction and the short range repulsion. 
On the other hand, the phase shift of the separable interaction is weakly
 attractive.
Because there is only one term in the rank I separable potential,
the contributions of attraction and repulsion can not be reproduced
 simultaneously.
In particular, the coupling strength $\lambda_{\pi}$ determined from the
 relation~(\ref{Vpi}) at low momentum region ($p=p'=0$) is too attractive, 
which therefore can not reproduce the repulsive behavior at higher $T_{lab}$.
However, one can fit the repulsive behavior at higher energies by changing the 
range parameter $m_{\pi}$ and the coupling constant $\lambda$.
For instance if we choose $m_{\pi}=\Lambda_{\pi}$$=$1300 MeV and
$\lambda=135\times 10^6$ MeV${}^2$, 
we can reproduce the phase shift at around $T_{lab}\approx$ 200 MeV as
 indicated by the dashed line of Fig.~\ref{fig:bonnandsp}-(c).  

\item

Fig.~\ref{fig:bonnandsp}-(d)  shows the phase shifts calculated
 from the potentials of the $\rho$ channel as functions of $T_{lab}$.
The phase shifts calculated from the two interactions look very different.
With $\lambda=-37.2\times 10^6$ MeV${}^2$ which is determined by
 Eq.~(\ref{Vpi}), the result of the separable potential is too attractive, 
such  that it generates one bound state and the phase shift starts from 180
degrees.  
In contrast, as shown in Fig.~\ref{fig:potential}-(b)   the
 $\rho$-exchange piece of the OBEP is attractive at long distances
 $r \gsim $ 0.6 fm and repulsive at middle and short distances
  $r \lsim$ 0.6 fm.
The attractive interaction here, however, is not very large due to the
 cancellation by the repulsive component.
Therefore the phase shift calculated from OBEP starts raising from 0 degree,
 turns to decrease at  $T_{lab}\sim 40$ MeV, and  change into
 repulsion at  $T_{lab}\sim 160$MeV.
Just as in the $\pi$ case, we can re-fit the strength of the separable potential
$\lambda_{\rho}$.
By reducing the strength by about factor 8, 
$\lambda=-4.52 \times 10^6$ MeV${}^2$, 
we obtain the phase shift of the separable potential as shown by the dashed
 line of Fig.~\ref{fig:bonnandsp}-(d), which looks rather close to the result
 of OBEP.

\end{itemize}

These results  show that it is difficult to reproduce the phase shifts of each
 terms of the one-boson-exchange potential separately by
 the rank I separable potential when we use the parameters determined in the
 long wave length limit. 
As explained above in detail, the separable potential can not be stronger than
a certain strength both for attractive and  repulsive cases if we do
not change the $m_b$ parameter.

 The fact that the separable potential can not be stronger than  a certain
 strength may be understood from Eq.~(\ref{tau}) where the factor
 $1/\lambda$ vanishes in the limit $|\lambda|\to\infty $.
Interestingly, in this limit there is no distinction between attractive
($\lambda\to -\infty $) and repulsive ($\lambda\to +\infty $) interactions.
In order to find the maximum strength of the separable potential, 
we plot in Fig.~\ref{fig:rehimh} the real and imaginary parts of $h(s)$
from which we can calculate the phase shift by using Eq.~(\ref{phaseshift}).
The result is shown in Fig.~\ref{fig:rehimh}.
The real part monotonically decreases from 0.276 GeV${}^{-2}$,
 while the imaginary part starts from 0, reaches the maximum value at some $s$
and turns to decrease monotonically.
This behavior resembles what is familiar in the non-relativistic scattering
theory where the phase shift varies from 0 to 180 degrees when there is one
 bound state.
In a relativistic theory, however, a naive argument in the non-relativistic
theory can not be applied, since in the large $s$ region particle production
may occur and the discussion within a fixed particle number can not be
 applied.
In the present separable potential model, the treatment will break down at
and beyond $s=4(M_N+m_b)^2$ where an unphysical pole of mass $m_b$ appears.
In our calculation of the phase shift, in order to determine the initial 
value $\delta(T_{lab}=0)$,
 for the attractive interaction we increased $\lambda$ gradually from a small
value and verified that  there is a jump from $\delta(T_{lab}=0)=0$ to
 $\delta(T_{lab}=0)=180$ degrees at certain strength of $\lambda$ only once.
Therefore, we conclude that  the maximum strength of the separable potential
in our method is what allows one bound state for an attractive interaction.
Similarly for the repulsive case, it is also possible to show that there is
 the maximum strength of the interaction if $m_b$ is fixed.

Now turning to the full result of the ${}^1S_0$ channel, as indicated in
 Fig.~\ref{fig:1s0}, the separable potential can
 reproduce rather well the result of the total nuclear force of OBEP
 when we take $\lambda=-0.294\times 10^6$ MeV${}^2$ and $m_b=224$ MeV
~\cite{Bondarenko:2002zz}.
The very strong attractive and repulsive forces of the $\sigma$- and
$\omega$-exchange potentials are largely canceled, 
yielding a rather mild nuclear force.
This is the reason that the separable potential for nuclear reaction
have been successful.

\begin{table}[h]
\centering
\caption{\label{bonn} \small Parameters of the OBEP from the 
Bonn potential~\cite{bonn}.}
\vspace*{0.5cm}
{\small 
\begin{tabular}{lccccc}
\hline
 &$m_b$(MeV)  & $g^2/4\pi$  & $\Lambda_b$(MeV)& $\lambda$(MeV${}^2$)
  \\
\hline
 $\sigma$ & 550 & 7.78 &2000&$-25.3\times10^6$\\
 $\rho$  & 769  &34.77 &1300&$-37.2\times10^6$\\
 $\omega$ & 783&20.0 &1500&$81.5\times10^6$\\
 $\pi$ & 138  &14.9  &1300&$-0.0189\times10^6$\\
\hline
\end{tabular}
}
\label{table1}
\end{table}
\begin{figure}[tbp]
\begin{minipage}{8cm}
\vspace*{0cm}
\centering
\includegraphics[width=8cm]{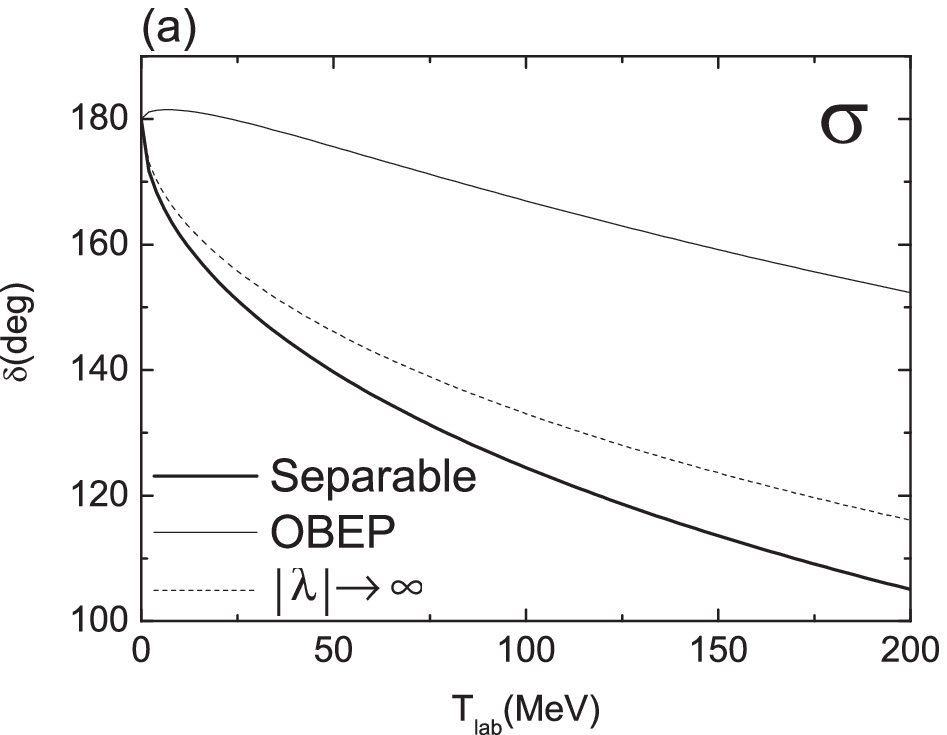}
\end{minipage}
\begin{minipage}{8cm}
\vspace*{0cm}
\centering
\includegraphics[width=8cm]{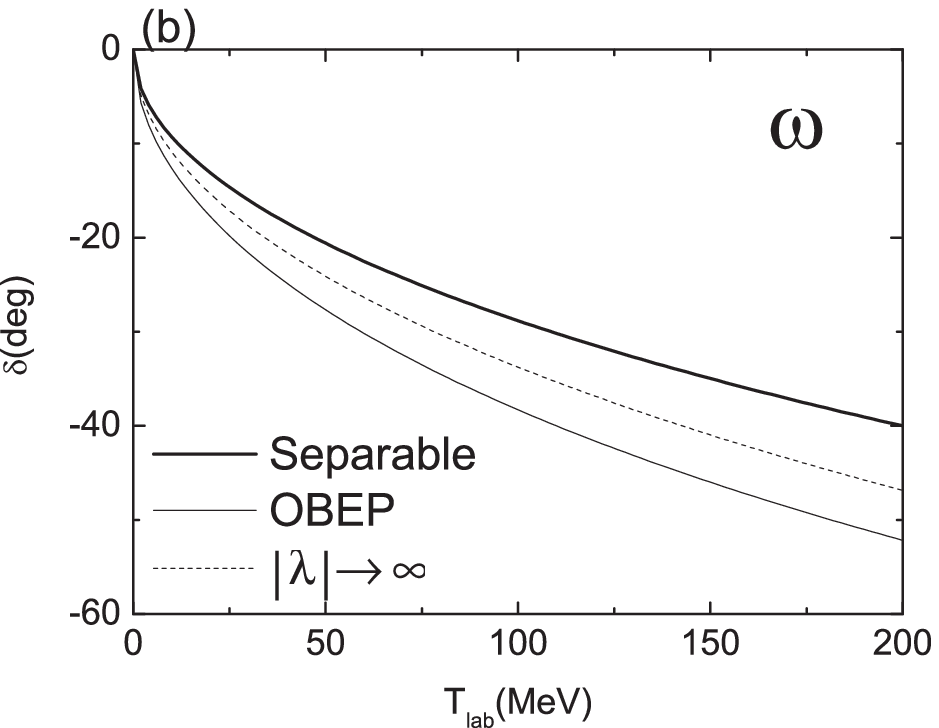}
\end{minipage}
\begin{minipage}{8cm}
\vspace*{0cm}
\centering
\includegraphics[width=8cm]{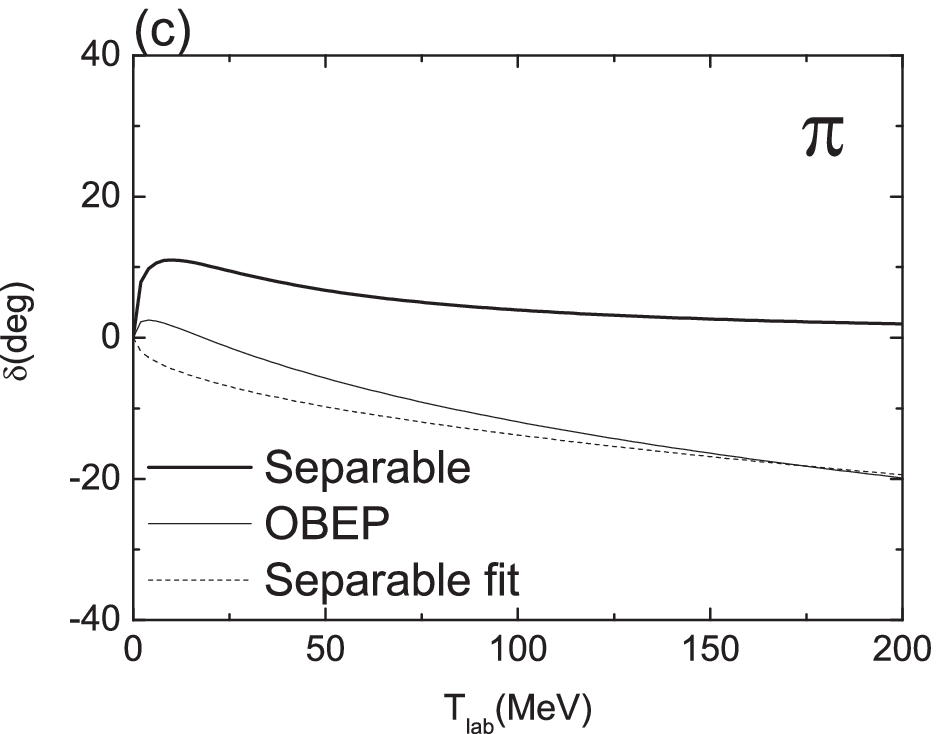}
\end{minipage}
\begin{minipage}{8cm}
\vspace*{0cm}
\centering
\includegraphics[width=8cm]{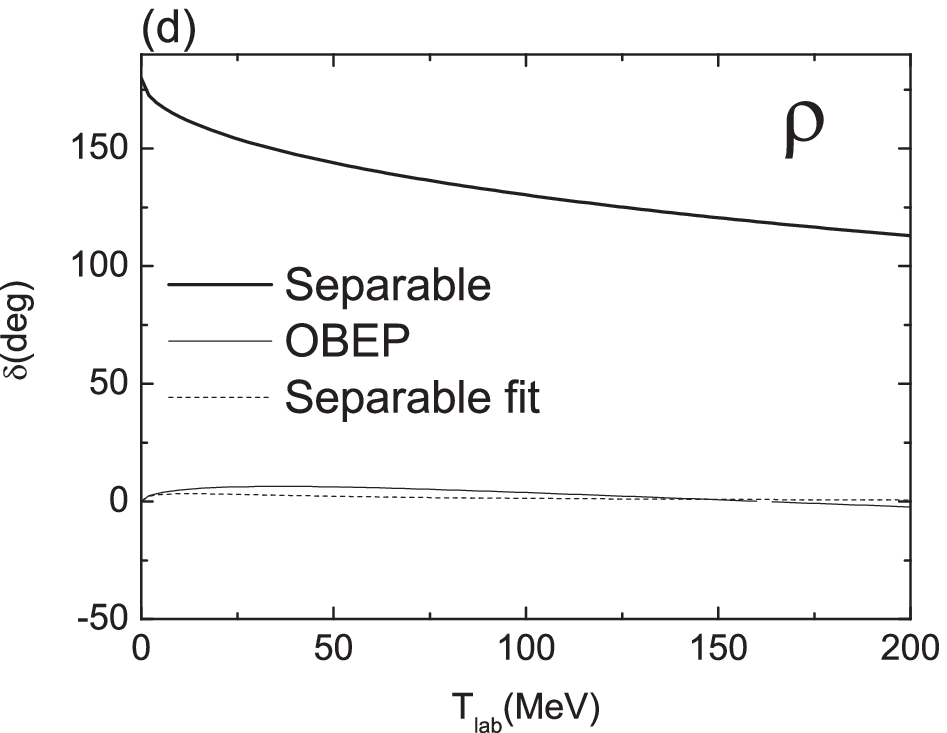}
\end{minipage}
\hspace{1cm}
\caption{\small Phase shifts of the ${}^1S_0$ channel calculated from the
 separable potentials (thick solid lines) and those from the OBEP
 (thin solid lines).
Dashed lines (a) and (b) represent the upper limit and the lower
 limit of the phase shift calculated from the separable potential
  with $|\lambda|\to \infty $.
The dashed line of (c)  represents  the phase shift fitted to the phase
shift of the OBEP around $T_{lab}\approx 200$MeV.
The dashed line of (d)  represents  the phase shift calculated by the
 separable potential which fits the
 phase shift of the OBEP around $T_{lab}\approx 0$ MeV.
 }
	  \label{fig:bonnandsp}
\end{figure}

	 \label{fig:omegafit}

\begin{figure}[tbp]  
\begin{minipage}{8cm}
\vspace*{0.0cm}
\centering
\includegraphics[width=8cm]{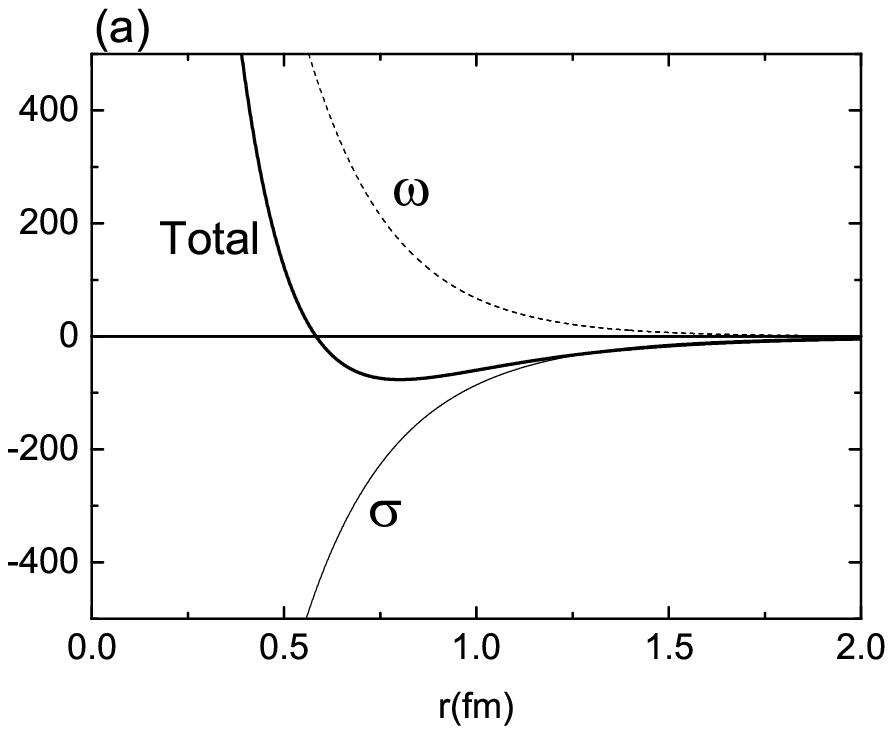}
\vspace{-1.0cm}
\end{minipage}
\begin{minipage}{8cm}
\centering
\includegraphics[width=8cm]{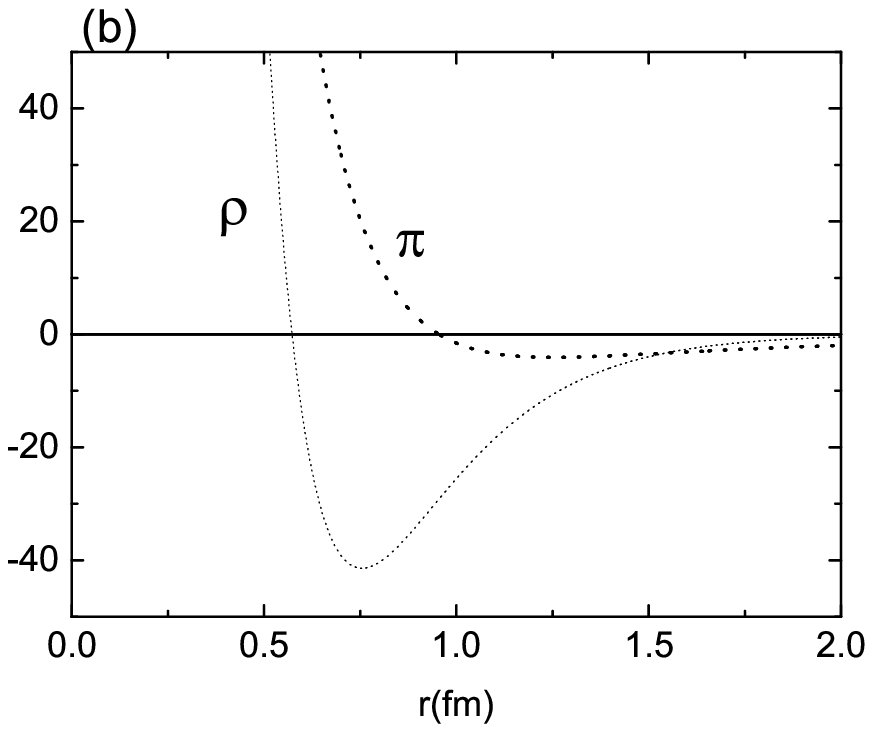}
\vspace{-1.0cm}
\end{minipage}
\caption{\small The separable contributions to the Bonn potential of the
 ${}^1S_0$ 
channel from  the various meson exchange terms as denoted by the labels.
(a): Thick solid, solid and thin dashed lines are for the total nuclear
 force, $\sigma$- and $\omega$-exchange potentials, respectively.
(b): Dashed and dotted lines are  $\pi$- and $\rho$-exchange
 potentials of the ${}^1S_0$ channel, respectively.}
	  \label{fig:potential}
\end{figure}

\begin{figure}
\begin{minipage}{16cm}
\centering
\includegraphics[origin=c,scale=1.20]{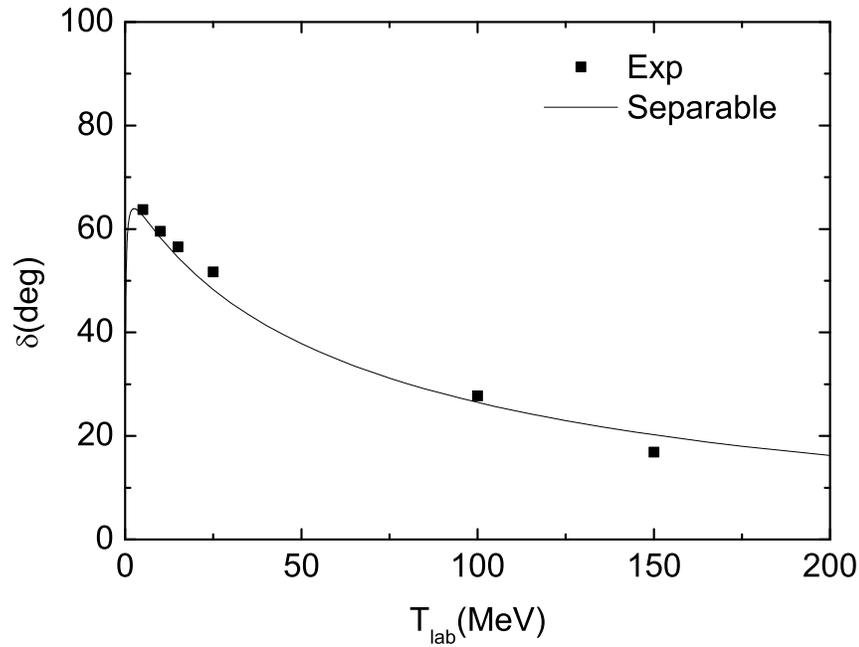}
\caption{\small 
The phase shifts of the ${}^1S_0$ channel calculated from the best fitted 
separable potential as a function of the kinetic energy in the laboratory
 frame as compared with the experimental data (data are taken using SAID
 program http://gwdac.phys.gwu.edu/).}
	 \label{fig:1s0}
\end{minipage}
\end{figure}

\begin{figure}
\begin{minipage}{16cm}
\centering
\includegraphics[origin=c,scale=1.20]{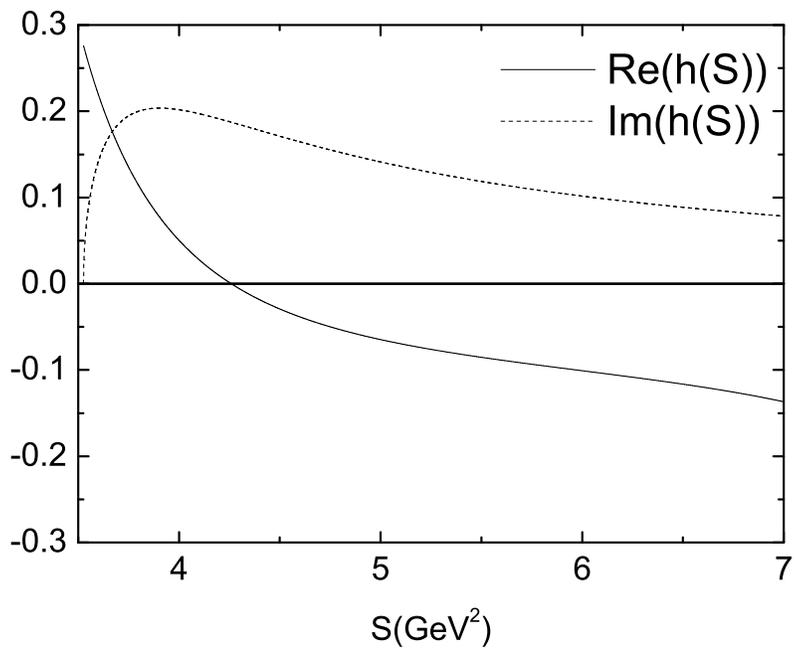}
\caption{\small 
The real and imaginary part of $h(s)$ when $m_b=500$(MeV) as a function of the
 total momentum square.}
	 \label{fig:rehimh}
\end{minipage}
\end{figure}

\section{Summary}

We have studied the relation between the rank I separable potential for
the covariant Bethe-Salpeter equation and the one-boson-exchange potential
 (OBEP).
Individual channels of $\sigma$-, $\omega$-, $\pi$- and $\rho$-exchanges were 
investigated separately. 
As a result, it turned out that the rank I separable potential could not
 reproduce the phase shift calculated from each component of the OBEP
when we use the parameters determined in the long wave length limit.
As for the $\sigma$ channel, where the potential is strongly attractive,
 we could not reproduce the phase shift of OBEP even if we take
 the limit 
  $\lambda\to-\infty$ and we change $m_b$ parameter.
Similarly as for the $\omega$ channel with strong attraction, the separable
 potential could not reproduce again the phase shift of OBEP
 even in the limit $\lambda\to\infty$. 
However we could reproduce the strong repulsion, if we change the $m_b$
 parameter.
These observations imply that the physical meaning of the mass parameters in
 the separable
potential and OBEP are different.
The mass parameter of the OBEP represents the interaction range of a local
 potential, while that of the separable potential could mimic, for instance,
the range of a non-local interaction.
The non-locality of the nuclear force is related to the structure of the
 nucleon at short ranges $r \lsim$ 0.5 fm~{\cite {ippan}}.
Concerning the $\pi$ and $\rho$ channels, where the  potential consists of
 attraction at long distances and repulsion at short distances,
the rank I separable potential could not reproduce the mixed nature of the 
interaction, although the interaction strengths are not as strong as the
 $\sigma$ and $\omega$ channels.
Despite the above fact, the rank I separable potential can reproduce the
 experimental data of the ${}^1S_0$ phase shift
up to the energy $T_{lab}\gsim 200$ MeV where a mild attractive interaction 
dominates.

These results show that the rank I separable potential is not suited to the
 description of very strong attraction.
For instance, phase shifts calculated from the separable potential can not 
become larger than 180 degrees, no matter how large the attraction coupling
constant takes.
Rather, the separable potential can describe relatively mild attraction and
 all repulsion.
In the realistic nuclear force, such a mild strength is obtained by the sum of
 the strongly attractive $\sigma$-exchange and the strongly repulsive
 $\omega$-exchange potentials.

In this work, we have shown that the separable potential works well 
for the two nucleon system if  parameters are chosen suitably,
although the decomposition into components of physical OBEP does not make
 sense. 
In  a sense, the different nature of the two potential should have been
 expected.
 The main purpose of the present paper was to see whether it is possible to 
make physical meaning of the
 separable potential in comparison with the OBEP by using a simple
 parameterization of one term of rank I.
In order to perform a good description of phenomena in the covariant
 Bethe-Salpeter formalism, we can introduce a higher rank form.
Such a work is now in progress,
where the use of the improved rank one ansatz and of higher
rank interactions are tested~{\cite {manabe}}.






 

\end{document}